# Roadmap for Gain-Bandwidth-Product Enhanced Photodetectors

V. J. Sorger[1,*] and R. Maiti[1]

[1]Department of Electrical and Computer Engineering, George Washington University, Washington, DC 20052, USA
*Email: sorger@gwu.edu

**Abstract:**
**Photodetectors are key optoelectronic building blocks performing the essential optical-to-electrical signal conversion, and unlike solar cells, operate at a specific wavelength and at high signal or sensory speeds. Towards achieving high detector performance, device physics, however, places a fundamental limit of the achievable detector sensitivity, such as responsivity and gain, when simultaneously aimed to increasing the detector's temporal response, speed, known as the gain-bandwidth product (GBP). While detector's GBP has been increasing in recent years, the average GBP is still relatively modest ($\sim 10^6$-$10^7$ Hz-A/W). Here we discuss photodetector performance limits and opportunities based on arguments from scaling length theory relating photocarrier channel length, mobility, electrical resistance with optical waveguide mode constrains. We show that short-channel detectors are synergistic with slot-waveguide approaches, and when combined, offer a high-degree of detector design synergy especially for the class of nanometer-thin materials. Indeed, we find that two dimensional material-based detectors are not limited by their low mobility and can, in principle, allow for 100 GHz fast response rates. However, contact resistance is still a challenge for such thin materials – a research topic that is still not addressed yet. An interim solution is to utilize heterojunction approaches for functionality separation. Nonetheless, atomistically- and nanometer-thin materials used in such next-generation scaling length theory based detectors also demand high material quality and monolithic integration strategies into photonic circuits including foundry-near processes. As it stands, this letter aims to guide the community if achieving the next generation photodetectors aiming for a performance target of GBP $\sim 10^{12}$ Hz-A/W.**

**Main Text**
Photodetectors deliver an integral function of optical-to-electrical (OE) conversion. Unlike solar cells performing this conversion spectrally broadband and temporally at steady-state, photodetectors and especially photoreceivers are typically designed for high responsivity (gain) at a target wavelength given by the absorbing materials' bandgap, and by fast-temporal dynamic response (3dB bandwidth, BW). Prominent material-spectrum mappings for detectors are GaN for UV (<400 nm) [1], Si for visible to NIR (400-1100 nm) [2], Ge/InGaAs for NIR to MIR (1-5 μm) [3] and HgCdTe for MIR to FIR (>5 μm) [4].

It is the aim of this letter to clarify i) fundamental photodetector tradeoffs, and ii) relate them to design choices, which thence govern detector performance. We show that this correlation is actually not arbitrary, and that there exist interdependencies between design-material combinations that yield a higher performance than others. The arguments laid out herein, are exemplary made on photonic integrated circuit (PIC)-based detectors, but their generality also holds for free-space coupled devices and applications.

The obtainable signal output from a detector (i.e. photocurrent and available photogain) and its operation response (speed, i.e. bandwidth, BW) are not interdependently optimizable, which can straightforwardly be understood by using the picture of the integration time of a photo collected signal; if the gain is high to increase the detectors sensitivity, then the response rate is slow, and vice versa. An example from daily life is the sluggish smart-phone camera response for taking photos at dim lighting conditions, where the digital signal processing circuit (DSP) increases the gain to enhance the signal-to-noise ratio (SNR), thus reducing the shutter speed.

Most of the photodetectors can be categorized broadly into two classes (**Fig. 1a**); those with high responsivity ($R$) (but low temporal bandwidth, speed) and vice versa, which relates to the fundamental trade-off gain-bandwidth product (GBP) being a figure-of-merit (FOM) for detectors. Since photodetector gain linearly scales with the responsivity ($R$) in units of (A/W), here, for discussion purposes, we use gain and responsivity interchangeably. Indeed, the iso-GBP lines show while devices cluster in either the upper left (low-BW/high-$R$) or lower right (high-BW/low-$R$) quadrant, both corresponding to a GBP ~ $10^5$-$10^6$ Hz-A/W (**Fig. 1a**). Naturally, detector performance and next-generation devices scale orthogonally to these two quadrants aiming for simultaneous high-speed and high $R$. It is the target to design and demonstrate such high-performance photodetectors that guides our motivation for the work presented herein (**Fig. 1a**).

Let us next analyze this fundamental GBP tradeoff in more detail with the aim to explore device paradigms that would allow to optimize the GBP beyond what is available with current technology; the photoconducting gain is given by the ratio of the photocarrier lifetime ($\tau_c$) to the photocarrier transit time ($\tau_t$), i.e. the drift time for photocarriers to reach the photocurrent-collecting electrodes. Meaning, if the carrier lifetime exceeds the time for their collection, charge accumulation (gain ≈ $\tau_c / \tau_t$) can occur. Given that the carrier lifetime in typically used detector semiconductors is about $O$(~ns), gains of about $O$(~$10^2$-$10^3$) can be obtained for transit times of 10's-100's of ps corresponding to medium-high mobility materials and micrometer source-drain distances. The transit time, on the other hand, is a function of the electrical transport properties (mobility, $\mu$) of the photoabsorbing material and the drift field ($V_{bias}$) via $\tau_t = L_{e/h} v_{drift}^{-1} = L_{e/h}^2 (\mu V_{bias})^{-1}$, where $L_{e/h}$ where is the distance of the electrodes collecting the electron-hole photocarriers (i.e. channel length, $L_{e/h}$).

Parametric BW-scaling of this mobility-limited transit time and RC-delay as a function of photodetector source-drain channel length, $L_{e/h}$, reveals several insights (**Fig. 1b**); a) if the channel length is several micrometer long mobilities of about 1,000 cm$^2$/Vs are required to even achieve 1 GHz fast detection, while RC is not a facto even for poor resistivities. b) in the limit for scaled (1 - 10 nm) short channel lengths even a poor mobility of 1 cm$^2$/Vs is not BW limiting up to about 100 GHz, provided at least a resistance of 1 kΩ can be guaranteed. c) a sweet spot for detectors that are aimed for speeds around 10's of GHz is found for channel lengths near 0.1 μm offering a high design window with relatively relaxed requirements ($\mu$ < 10 cm$^2$/Vs and R < 10 kΩ). As an interim conclusion we can summarize that if high-speed devices (>100 GHz) are desired, then transistor-short channel lengths (~10 nm) are required, since even high-mobility (~10,000 cm$^2$/Vs) materials cannot reach these BW for micrometer wide contacts use in SiO$_2$ or Lithium-niobite (LN)-based waveguides. For III-V and silicon-based waveguides a channel length of 500 nm is already challenging, if plasmonic losses from the contacts are to be avoided. Thus, for our exemplary 100 GHz goal, mobilities of 1000 are required which is not possible for low-resistance (highly-doped contacts). Thus, it seems an interesting design option to accept possible optical losses from metal-optics and design 10's nm narrow slot-waveguides, which can achieve 100 GHz devices with both modest mobilities ~5 cm$^2$/Vs and resistances ~5 kΩ). Such a material is, for instance, the class of transition metal dichalcogenides (TMDC). In fact, one can show that borrowing device scaling laws from electronics, that 'flatness' of the detector material is a key requirement, as discussed further below.

However, since the gain is proportional to both the mobility and the carrier lifetime, the former can limit the bandwidth of the detector such as for poor mobilities known form TMDCs [29]. Yet, there are several reports on ultrahigh gain in TMDCs based photodetector where the underlying mechanism is mainly photogating effect instead of multicarrier generation or avalanche amplification [20, 21]. In the photogating effect, the charge carrier are trapped in surface impurity states, thus resulting in a DC-slow response time (1-10 s), too slow for many data communication and signal processing/computing, but possibly sufficient for environmental sensing applications [17, 18, 20, 21, 36]. Therefore for long-channel detectors, in order to achieve high GBP, focus should be on improving carrier mobility rather than prolonging carrier lifetime, while the inverse is required (i.e. focus on improving contact resistance) for short-channel detectors (**Fig. 1b**).

Borrowing device scaling concepts from electronics, the *scale-length theory* (SLT) shows that electric field-based device dimensions scale with a nominal factor (σ) such as in FETs both the channel thickness and length [5]. Applying this dimensional scaling concept to the GBP of photodetectors, we find that it scales as GBP ~ $σ^{-2}$ ~ $L_{e/h}^{-2}$, which suggests that reducing the collection distance rather sensitively (squared) improves detector performance. Hence the question arises, as of why prevailing integrated detector designs do not make use of this scaling performance gains?

The answer can be found in the inability of monolithic material integration to adhere to the SLT scaling; starting our GBP optimization analysis discussion with optical constrains, utilizing this $σ^{-2}$ scaling law, urges device designers to utilize nanometer (or 10's of nanometer) short-channel (contact distances) to optimized GBP. This, however, requires a simultaneous 'squeezing' of the optical mode to the same order (10's nm compact), which can be realized with nano-optical light-matter-interaction enhancements such as offered by plasmonics [6] or dielectric discontinuities [7]. SLT applied to detectors simultaneously demands nanometer thin absorbing materials, which is further enforced to ensure a high optical mode overlap (Γ) [8], similar to quantum-well lasers [9]. Realizing such a nanometer-thin absorbing material at high crystalline quality, however, is challenging to be provided by III-VI and IV (e.g. Ge) materials when these materials are heterogeneously integrated in Si or SiN photonic platforms due to the parasitic high defect density during patterning (e.g. reactive ion etching) [10,11]. However, a SLT-scaled device detector is synergistic with two-dimensional (2D) materials, as briefly mentioned above and further discussed next (**Fig. 1**).

As a possible high GBP-design example, a rather short the plasmonic slot atop a silicon on insulator (SOI) waveguide, where $L_{e/h} = W_{slot}$ (slot-width ~10 nm), supports a gap plasmon with an electric field polarization being in-plane with such a nanometer thin material [12, 13]. The detector's external quantum efficiency is given by: EQE = IQE x (1-Reflection-Transmission), where the internal quantum efficiency (IQE) is proportional to the overlap matrix element in the absorption process (typically ~80's - 90's% for band-to-band absorption [14]). The second term relates to the amount of light absorbed by the device, which can be optimized by anti-reflection coatings for free-space coupled devices. In integrated devices, however, this relates to the mode overlap factor ($\Gamma$), i.e. the amount of light coupled to the materials. Since a detector's photo responsivity is proportional to the EQE, and hence to GBP, ensuring a high optical absorption is key. With 2D materials featuring a rather high intrinsic absorption coefficient (~$10^5$ cm$^{-1}$), they are an interesting SLT-synergistic material option; furthermore, their atomistically-thin flatness reduces coulomb scattering effects, thus allowing for a high oscillator strength [15] signified by a high optical index $Re\{n\} > 5$ in the visible and NIR bands for TMDCs [16]. Furthermore, such a ~10 nm short electrical channel, also comes with (near) ballistic electric transport. Hence, a high drift mobility would actually not improve the transit time which is <1 ps, for $V_{bias} = 1$ V and $\mu_{TMDC}$ = 1-10 cm$^2$/Vs, and the low mobility of 2D materials (aside Graphene) is not a limit of a SLT-scaled high-GBP 2D material-based detector.

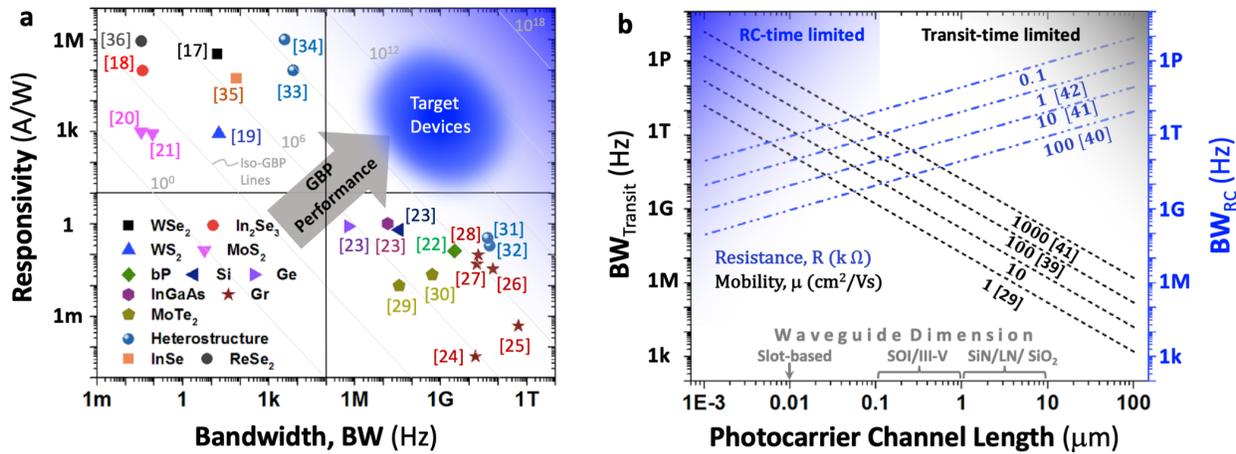

**Fig.1. Design strategies for Engineering Gain-Bandwidth-Product (GBP) Photodetectors.** a) Photodetector performance cluster into two quadrants ; high responsivity yet low speed (bandwidth, BW), and vise versa. Target detectors with a GBP of ~$10^{12}$ scale orthogonally (upper-right quadrant) to the iso-GBP lines. The values are taken from [17-36]. b) Plot of bandwidth vs. photocarrier-collecting electrical channel length showing two regimes where the transit time-limited bandwidth (BW$_{transit}$) is dominant for long-channel detectors (gray shade), whereas for sub 100 nm channel lengths, the bandwidth is limited by RC time (blue shade). BW$_{transit}$ is estimated for four different mobility values (1 - 1000 cm$^2$/Vs) for a $V_{sd} = 1$V taken from ref [37- 39]. The RC time limited bandwidth (BW$_{RC}$) is estimated for four different resistance values starting from 0.1–100 k$\Omega$ [40-42], where, the electrical capacitance is determined by a parallel-plate model (fringe fields ignored) for a lateral junction. A finding of this parametric study is that for a (arbitrarily set) target speed of 100 GHz; i) micrometer long channel-based photodetectors required a very high mobility (<$10^4$ cm$^2$/Vs) with a relaxed contact resistance, while ii) 10's nm short-channel detectors are only resistance limited, since (near) ballistic transport is given even for poor mobilities (<10 cm$^2$/Vs). Such short-channel performance-detectors, however demand optical mode squeezing to adhere to device *scale-length theory* (SLT) rules which can be achieved with slot-waveguide designs.

However, improvement a material's mobility does relax length-scaling requirements (const. = GBP → μ ~ $(L_{e/h})^2$ (**Fig. 1b**). For instance, some recent undoped TMDC's have shown mobilities of $10^3$ cm$^2$/Vs [37,38], which allows increasing $L_{e/h}$ from 10's to 100's nm for high-speed detectors. Channel-length scaling, however, also impacts the detector's dark current ($I_{dark}$); generally speaking, lowering $I_{dark}$ increases the detectors dynamic range, and improves the noise-equivalent-power (NEP), which is the required optical power to produce a SNR of unity at 1 Hz signal rate. Indeed, we [29] and others [30] have shown, that a TMDCs-based detector is able to operate at 1-10 nA dark currents, while a scaled slot detector has an about 10x higher $I_{dark}$. Compared to graphene, TMDCs feature a 10-1000x lower dark current due to their 0.5 - 2 eV wide bandgap, whilst a tiny energy (optical or thermal) will initiate intra-band transitions in graphene. Another possibility for reducing the dark current, is to operate the detector in photovoltaic mode, namely at zero bias voltage, instead of the often used 1-2 V, however also give up any gain enhancement. Such a detector can be realized with built-in P/N junction for charge separation polarity; for instance, surveying the literature, points to a potential of 10-100x lower $I_{dark}$ [41,42].

However, 2D materials also have challenges in designing detector (e.g. high resistance), however as discussed before, their low-mobility is not one of them, provided SLT short $L_{e/h}$ approaches are used. Yet, total resistance of the device comprises of contact resistance ($R_c$) and channel resistance ($R_{ch}$), is the key to achieve high-performance. Since $R_{ch}$ quickly decreases with the channel length scaling, $R_c$ becomes a dominant factor in the total resistance for sub-100 nm channel lengths. For instance, the high contact resistance in kΩ to even MΩ challenges the device response time limiting RC-delay [22] (**Fig. 1b**). A solution to this is to design hetero-junctions, where function-material separation can be part of the design process; for instance, a high bandgap TMDC placed atop a SOI waveguide performs the OE conversion, whilst the photocarriers are transferred into graphene layers which acts as an electrical conducting channel or as a ~50Ω contact resistance to the metal contact pads [30]. Several contact engineering techniques including edge contact geometry, 1T phase contacts, hBN substrate and graphene have been implemented in individual devices, but a scalable route for uniformly realizing these device concepts at the wafer-scale has not yet to be established [43,44].

Taking the above discussed high-GBP design options into account on high gain and responsivity via short transit-times, low dark-current via zero-bias using junctions, heterojunction devices for low RC, we can estimate a speed of ~10-100 GHz (limited by RC) (**Fig. 1b**), and a responsivity of 1-100 A/W (assuming a carrier lifetime ~$\boldsymbol{O}$(ns)), thus enabling a GBP ~ $10^{12}$ Hz-A/W (blue region, **Fig. 1a**), which would be an about 1-2 orders orders of magnitude improvement over record GBP devices, and about 4-5 orders improvement compared to the average detector performance (**Fig. 1a**). With an mobility-improved and hence L-relaxed $L_{e/h}$ of, for instance 100 nm, the resulting $I_{dark}$ can approach $\boldsymbol{O}$(100) pA for an NEP of ~1 pW/Hz$^{0.5}$ rivaling state-of-the-art detectors [23].

Interestingly, 2D materials have just recently reached foundry maturity such as demonstrated by IMEC [45] or TMSC [46], for example. This recency bears hope for further material improvements with R&D economy-of-scale testing in the near future, and monolithic growth on SOI-oxide surfaces is possible, yet quality levels such a s defect density still required attention. While CVD such processes are actively being developed to date, transfer techniques for 2D materials have successfully addressed issues such as deterministic placement [47] and single-flake transfer showing vanishing cross-contamination, past transfer [48]. For instance, micro-stamping techniques allow for rapid-prototyping devices benign to PIC heterogeneous 2D material integration enabled by the high spatial selectivity of the micro-stamper [49].

Furthermore, 2D materials offer a higher degree of strain-material adaptation given added range of design flexibility, which we termed 'strainoptronics'; for instance, straining (tensile or compressive) dramatically (100's of meV) shifts the band edge of 2D materials leading to enhanced absorption so as responsivity for 2D materials based photodetector [29]. The understanding of the modulation of optical and electronic properties of these materials are key to design improved integrated photonic devices [50-52]. Techniques such as Kelvin probe force microscopy (KPFM) allow for local strain mapping which allows for nanometer-resolution strain and hence bandgap maps thus allowing to align the photoabsorbing 2D material with photonic structures [53] such as waveguides, or optical cavities precisely [29]. Indeed, the large amount of strain that 2D materials can sustain before rupture occurs along with their large elastic modulus [52], which allows reducing the bandgap of, e.g. $MoTe_2$, from 1.04 eV down to 0.80 eV, thus enabling efficient photo conversion at the telecommunication relevant wavelengths of 1550 nm [29]. Indeed, 2D materials offer unique properties for number of opto-electronic applications beyond detectors such as for nanoscale emitters [54] or electro-optical modulators [55,56] in applications ranging from data-communication [57], over neural networks [58], to bio-photomodulation [59].

In conclusion, like other opto-electronic fundamental building blocks, the device physics of photodetectors dictates fundamental trade-offs between detector sensitivity such as available gain, responsivity and hence noise-equivalent power on the one side, and bandwidth, speed, on the other. Here we discussed photodetector gain-bandwidth-product (GBP) performance in the context of device channel scaling, electrical and opto-electronic performance constrains. We show that state-of-the-art detectors are either optimized for gain or speed showing an averaged GBP of $10^6$ to $10^8$ Hz-A/W (using responsivity for gain due to linear proportionality amongst them). Further, we discuss strategies for achieving next-generation GBP of $\sim 10^{12}$ Hz-A/W following scaling length theory. For instance, we show that for achieving 100 GHz fast detectors micrometer-wide separated photocarrier-collecting source/drain contacts require exceedingly-high mobilities ($\sim 10^4$ $cm^2/Vs$) which is unrealistic in simultaneously achieving to low contact resistances required to reduce RC-delay. A seemingly more elegant, and PIC-density benign approach of achieving high GBP detectors, is to scale the channel length into the sub 100 nanometer regime, where transport is near ballistic even for low-mobility ($\sim 1-10$ $cm^2/Vs$) materials. Interestingly, we find that the scaling length theory and plasmonic slot-waveguide approaches offer a high-degree of detector design synergies especially for flat materials. Such 2D materials may be a solution for such next generation high GBP devices based on channel length scaling where their low mobility does not limit performance. However, several issues such as scalability, contact resistance, doping and mobility still need to be standardized prior to replacing existing technology. Therefore, in order to realize the fullest potential of such device physics-benign emerging materials, constrains from foundry-based mass productions and chip integration should be co-developed to ensure satisfactory and reliable performance.